\documentclass[a4paper,twocolumn,11pt,accepted=2022-03-15]{quantumarticle}
\pdfoutput=1
\usepackage[utf8]{inputenc}
\usepackage[english]{babel}
\usepackage[T1]{fontenc}
\usepackage{amsmath}
\usepackage{hyperref}

\usepackage{tikz}
\usepackage{lipsum}

\newcommand{\nbar}{\overline{n}}
\newcommand{\partialD}[2]{\frac{\partial #1}{ \partial #2}}
\newcommand{\tr}[1]{\mathrm{tr}\left[#1\right]}

\begin{document}

\title{Entropy production and fluctuation theorems in a continuously monitored optical cavity at zero temperature}

\author{M. J. Kewming}
\affiliation{School of Physics, Trinity College Dublin, College Green, Dublin 2, Ireland}
\email{kewmingm@tcd.ie}
\author{S. Shrapnel}%
\affiliation{Centre for Engineered Quantum Systems, School of Mathematics and Physics, University of Queensland, QLD 4072 Australia}

\maketitle

\begin{abstract}
Fluctuation theorems allow one to make generalised statements about the behaviour of thermodynamic quantities in systems that are driven far from thermal equilibrium.
In this article we use Crooks' fluctuation theorem to understand the entropy production of a continuously measured, zero temperature quantum system; namely an optical cavity measured via homodyne detection.
At zero temperature, if one uses the classical definition of inverse temperature $\beta$, then the entropy production becomes divergent.
Our analysis shows that the entropy production can be well defined at zero temperature by considering the entropy produced in the measurement record leading to an effective inverse temperature $\beta_{\rm eff}$ which does not diverge.
We link this result to the Cram\'er-Rao inequality and show that the product of the Fisher information of the work distribution with the entropy production is bounded below by half of the square of the effective inverse temperature $\beta_{\rm eff}$.  
This inequality indicates that there is a minimal amount of entropy production that is paid to acquire information about the work done to a quantum system driven far from equilibrium.
\end{abstract}

\section{Introduction}
The laws of thermodynamics provide a tremendously useful set of scientific tools. 
They tell us which natural physical processes are possible, show us that our machines have efficiency limits, and arguably form a basis for our understanding of the arrow of time. 
In the preceding decades the scope of thermodynamics has enlarged to include nonequilibrium systems, quantum systems, and cosmological systems such as black holes. Thermodynamics is a truly universal theory.
In particular, quantum thermodynamics is a rapidly evolving discipline bringing together concepts from quantum information,
many-body physics, and nonequilibrium thermodynamics \cite{goold_role_2016, vinjanampathy_quantum_2016}.

For small thermalising systems with few interacting degrees of freedom, fluctuations will dominate and the thermodynamic quantities must be characterised by stochastic random variables \cite{seifert_stochastic_2012}. 
Investigations into these fluctuations has established a new understanding regarding constraints on entropy production and the second law of thermodynamics \cite{jarzynski_nonequilibrium_1997, crooks_entropy_1999, hatano_steady-state_2001, seifert_entropy_2005}. 
These seminal results, which are now colloquially known as fluctuation theorems (FTs), have established a remarkable insight; fluctuations in microscopic systems that are comparable in scale to the system have a specific mathematical structure determined entirely by the entropy production $\Sigma$ \cite{horowitz_thermodynamic_2020}. Thus classical FTs have significant implications for understanding energy exchange in biology, nano-engineering, computation and communications. 

Another pertinent line of research stemming from this body of work is the emerging field of quantum stochastic thermodynamics (QST) \cite{strasberg_operational_2019, strasberg_non-markovianity_2019}.
Quantum systems are unique in the sense that the measurement process is intrinsically connected to their stochastic behaviour \cite{davies_quantum_1969, gardiner_quantum_2004, wiseman_quantum_2009}. 
In QST, researchers seek to understand the nature of heat $Q$, work $W$, and entropy production $\Sigma$, and their associated probability distributions---$P(W), P(Q), P(\Sigma) $---when subject to continuous measurement.
A natural question to ask is whether continuously measured quantum systems satisfy the classical FTs, or whether these theorems need to be modified in accordance with experimental findings.
Given the stochastic nature of quantum measurement, it is interesting to consider how FTs could be incorporated into the frameworks of open and closed quantum systems \cite{mukamel_quantum_2003, popescu_entanglement_2006, esposito_nonequilibrium_2009, landi_irreversible_2020, campisi_colloquium_2011,  hasegawa_thermodynamic_2021, miller_joint_2021, Stefano_noequilibrium_2018}.
Seminal results have shown that quantum measurements produce entropy \cite{horowitz_quantum-trajectory_2012, horowitz_entropy_2013, alonso_thermodynamics_2016, elouard_role_2017, manikandan_fluctuation_2019, belenchia_entropy_2020, naghiloo_heat_2020}, and that perfect measurements cannot be performed on quantum systems using finite resources \cite{guryanova_ideal_2020}.
Further work has shown that measurement can be utilised as a resource in quantum engines for information, work extraction, and battery charging \cite{elouard_extracting_2017, naghiloo_information_2018, monsel_energetic_2020, mitchison_charging_2021}.
Importantly, the fluctuations induced by the measurement process satisfy several thermodynamic uncertainty relations (TURs) \cite{hasegawa_quantum_2020, hasegawa_thermodynamic_2021} which impose strict restrictions on the fluctuations of thermodynamic current \cite{barato_thermodynamic_2015, pietzonka_universal_2016, pietzonka_universal_2018}.
This is particularly important when considering the limitations of precision from quantum measurement \cite{van_vu_thermodynamic_2020} and quantum or classical feedback \cite{vu_uncertainty_2020, potts_thermodynamic_2019, debiossac_thermodynamics_2020, cao_thermodynamics_2009, horowitz_nonequilibrium_2010, Gong_quantum_2016, murashita_fluctuation_2017}.
For a recent comprehensive review of stochastic quantum thermodynamics for continuously measured systems, please see \cite{manzano_quantum_2021}.

Despite this growing body of work, it is not yet clear how to extend FTs to the zero temperature setting, $T{\rightarrow} 0$, where the inverse temperature $\beta{=}1/k_{B}T$ diverges. 
One could argue that the zero temperature limit is unphysical---citing, for example, the third law of thermodynamics, or showing that this limit requires infinite resources or infinite time \cite{masanes_general_2017}.
Despite this argument, the zero temperature limit is nonetheless a very good approximation for optical cavities, and is of significant practical utility in theoretical modelling of quantum optical systems. 
As such, it is important to understand how entropy production can extended to the zero temperature limit, particularly for quantum systems undergoing continuous measurement.

One proposed solution is to use the Wigner entropy, which relates $\beta$ to the zero point energy, ensuring $\Sigma$ remains well defined at zero temperature \cite{santos_wigner_2017, belenchia_entropy_2020}. Wigner entropy is defined in terms of a quasi-probability distribution over phase-space, and for pure states can be non-zero \cite{adesso_measuring_2012}). Interestingly, Wigner entropy has been shown to satisfy an integral fluctuation theorem, thus extending its application to stochastic systems driven far from equilibrium. However, what is lacking from the Wigner entropy description is its relationship to stochastic fluctuations arising in the measurement record from continuous measurement, as is typically described in QST.

In this paper we take a new approach and show that both entropy production, $\Sigma$, and FTs can be well defined for continuously measured quantum systems at zero temperature by considering the entropy of the measurement record. 
This is done in the context of a single mode optical cavity, driven from an initial vacuum state to a nonequilibrium steady-state, and monitored continuously via homodyne detection (similar to the models considered in \cite{horowitz_quantum-trajectory_2012, santos_wigner_2017, belenchia_entropy_2020}).
Assuming that the optical cavity is driven from an initial thermal equilibrium state to a non-equilibrium steady-state, then $\Sigma$ is related to the difference between work done $W$ during the process and the Helmholtz free energy of the system $\Delta F$ \cite{crooks_entropy_1999}:
\begin{equation}
\label{eq:ep_crooks}
    \Sigma = \beta (W - \Delta F).
\end{equation}
We show that if one defines $\Sigma$ using the classical Shannon entropy of the measurement record used to infer the \emph{estimated} work $W$, then to satisfy Eq.\,(\ref{eq:ep_crooks}), one must define an \emph{effective} inverse temperature, $\beta_{\mathrm{eff}}$, which is determined by the measurement efficiency $\eta$, the mean photon number of the cavity $\nbar$, and the average energy in the zero temperature bath. 
We further show that this definition produces results that agree with those derived using Wigner entropy \cite{santos_wigner_2017, belenchia_entropy_2020}.
Interestingly, while $\beta_{\mathrm{eff}}$ differs from the classical inverse temperature, it agrees in the high-temperature limit when one has access to perfect measurements, corresponding to $\beta_{\mathrm{eff}} \rightarrow 1/k_{B}T$.


As a final consideration in this article, we seek to understand the relationship between the entropy production $\Sigma$ and what we can infer about the average work from the measured signal. A natural information measure to consider is the classical Fisher information (FI) of the average measured work $I(W)$. We compute the FI from the observed work statistics and relate it to the average entropy produced $\Sigma$. We derive an inequality that bounds the product of $\Sigma$ and the $I(W)$ below by half the effective temperature squared $\beta_{\rm eff}^{2}/2$. Thus, for a given FI, only a given region of possible $\Sigma$ values are permitted. This highlights the direct trade-off between information gain and entropy production in such continuously measured systems.

\section{System Dynamics and Thermodynamics}
\label{sec:System Dynamics and Thermodynamics}
We begin by considering a single mode cavity, with frequency $\omega$ and annihilation (creation) operators $\hat{a}\, (\hat{a}^{\dagger})$.
The cavity is coherently driven by an AC driving field, on resonance with a pump-frequency $\omega$ along the phase quadrature $\hat{q}$, with a maximum amplitude $g$.
The Hamiltonian describing this cavity system is
\begin{equation}
\label{eq:Hamiltonian}
    \hat{H}(t) = \hbar \omega \hat{a}^{\dagger}\hat{a} + g \sin(\omega t) \lambda(t)\hat{q}\,,
\end{equation}
where the single mode phase quadratures are $\hat{q} {=} \sqrt{\hbar/2\omega}(\hat{a} + \hat{a}^{\dagger})$ and $\hat{p} {=} i\sqrt{\hbar\omega/2}(\hat{a} - \hat{a}^{\dagger})$,
and satisfy $\left[ \hat{q}, \hat{p}\right]=i\hbar$.
The $\sin(\omega t)$ term ensures that our optical field will physically couple to our cavity on resonance. 
The power of the drive is time dependent $\lambda(t)$; initially equal to zero $\lambda(0) {=} 0$ (i.e. the drive is turned off) and then ramps up to $\lambda(\tau) {=} 1$ (the drive is turned on).
Unsurprisingly, when the drive is on $\lambda(t)\neq0$, the AC drive does work $W$ on the internal cavity state $\hat{\rho}$ due to the continuous oscillatory term.

We will further assume that our system is weakly coupled to a thermal reservoir of Bosons which evolves under the Born-Markov assumption.
The strength of this coupling is $\gamma$.
We characterise the mean photon number of the bath via the Bose-Einstein distribution ${\nbar{=}(\exp(\mu) - 1)^{-1}}$, where $\mu {=} \hbar\omega/k_{B}T$. 
Notably $\nbar$ will approach zero when $T{\rightarrow}0$ indicating the environment is a vacuum.
We can describe the evolution of the internal cavity state via the local Lindblad master equation, which in the Schr\"{o}dinger picture
\begin{align}
    \frac{d\hat{\rho}}{dt} &=-\frac{i}{\hbar} [\hat{H}(t), \hat{\rho}] + \gamma_{-}\mathcal{D}[\hat{a}]\hat{\rho} + \gamma_{+}  \mathcal{D}[\hat{a}^{\dagger}]\hat{\rho}\,,
\end{align}
where we have $\mathcal{D}[\hat{a}]\hat{\rho} {= \hat{a}\hat{\rho} \hat{a}^{\dagger} - (\hat{a}^{\dagger}\hat{\rho} +\hat{\rho} \hat{a}^{\dagger})/2} $ as the standard Lindblad dissipator, and $\gamma_{-} {= \gamma(\nbar + 1)}$ and $\gamma_{+}{=}\gamma \nbar$ are the emission/absorption rates respectively.
Initially, the internal cavity state is described by a thermal Gibbs state $\hat{\rho}_{i} {=} e^{-\mu \hat{a}^{\dagger}\hat{a}}/\mathcal{Z}$ where $\mathcal{Z}{=}\mathrm{tr}(e^{-\mu \hat{a}^{\dagger}\hat{a}})$ is the Partition function.

From the perspective of quantum thermodynamics and the local Lindlad master equation, the average power $\langle dW \rangle$ from the drive in the weak coupling limit is \cite{vinjanampathy_quantum_2016}
\begin{equation}
\label{eq:work}
    \langle dW \rangle = \left< \partialD{\hat{H}(t)}{t}\right> \approx g \omega  \lambda(t) \langle \hat{q}(t) \rangle\,,
\end{equation}
where we have only included the dominant term proportional to $\omega$.
The average work done over the time $t\in (0,\tau]$ is given by the integral $\langle W \rangle = \int_{0}^{\tau}\langle dW \rangle dt$.
If we substitute the Hamiltonian Eq.\,(\ref{eq:Hamiltonian}) into this expression, and transform into the interaction picture under the rotating wave approximation, we obtain an expression for the average work
\begin{equation}
\label{eq:work2}
    \langle W \rangle =g \omega  \int_{0}^{\tau}dt \lambda(t) \langle \hat{q}(t) \rangle \,.
\end{equation}
Thus we can directly infer $W$ by measuring the evolution in $\hat{q}$ in this frame. 

\begin{figure}[t]
    \centering
    \includegraphics[width=0.99\columnwidth]{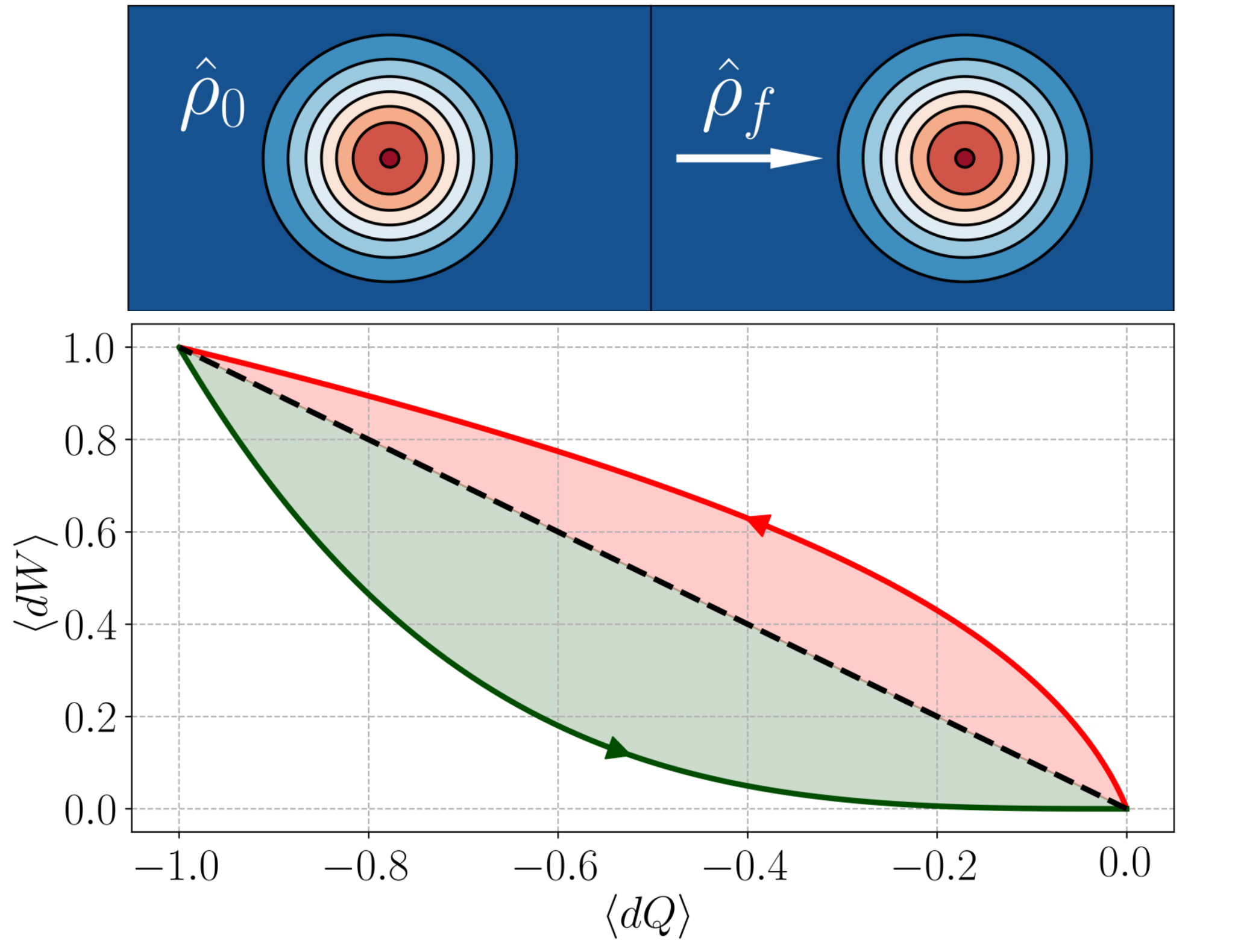}
    \caption{(Top) The Wigner function for the initial and final states. The initial state is a thermal state centred at the origin, whereas the final state corresponds to a displaced thermal state. (Bottom) If we plot the power $\langle dW \rangle$ against the dissipated heat $\langle dQ \rangle$ for the forward (Red) and reverse (Green) paths. When the path intersects the black dashed line, the system is in a steady-state. At the origin, this corresponds to the equilibrium steady-state, where all other cases when $\langle dW \rangle = - \langle dQ \rangle \neq 0$ correspond to NESS.  The simulation parameters are $\nbar = 1$, $\gamma=2$, and $g =1$. Here we will assume the power ramp of the drive $\lambda(t) = (\exp(-\sigma(t - t_{0}) + 1)^{-1}$ follows a sigmoid profile where $\sigma$ controls the steepness of the ramp and $t_{0}=\tau/2$ corresponds the centre of the ramp. }
    \label{fig:NESS}
\end{figure}

The dynamics in this picture are therefore quite straightforward; the system is initially in a Gibbs state, centred at the origin of phase space which is then driven over a period of time $\tau$, to a displaced Gibbs state---in the rotating frame with $\lambda(t) {=}1$--- with a mean steady-state displacement $\langle \hat{q} \rangle_{ss} =2g/\omega \gamma$ as depicted in Figure\,\ref{fig:NESS}.
This displaced thermal state corresponds to a nonequilibrium steady-state (NESS) given that the average power added $\langle d W \rangle$ is equal the average dissipated heat current $\langle dQ \rangle$---in this case, due to photons leaking out of the cavity per unit time and is computed via
\begin{equation}
    \langle dQ \rangle = \frac{d \langle \hat{H}(t)\rangle}{dt} \approx \hbar \omega \gamma(\nbar - \langle \hat{n}\rangle)\,,
\end{equation}
again where we have only included the dominant term proportional to $\omega$.
This means that the average change in internal energy $\langle dU \rangle =\langle dW\rangle + \langle dQ\rangle$, is zero.
Thus the non equilibrium steady-state occurs when $\langle dW \rangle = - \langle dQ \rangle$.
If we plot these two processes against one another, we can easily see where the steady-states emerge and the energy cycle traced out by this system in Figure\,\ref{fig:NESS}.

Lastly, we can compute the change in Helmholtz free energy $\Delta F$, which is defined as the difference between the change in internal energy $\Delta U$ and the change in the Von-Neumann entropy in the system $\Delta S$
\begin{equation}
    \Delta F = \Delta U - T \Delta S\,,
\end{equation}
where $T$ is the temperature system. 
It is easy to show that the $\Delta S{=} 0$ since $\hat{\rho}_{i}$ and $\hat{\rho}_{f}$ are both thermal, and displaced thermal states respectively.
Therefore one can readily show that the change in Helmholtz free energy between $\lambda(0) =0$ and $\lambda(\tau)=1$ as
\begin{equation}
    \Delta F  = -\hbar \omega \left(\langle \hat{n} \rangle_{f} -\langle \hat{n} \rangle_{0}  \right) = -\frac{2\hbar \omega g^{2}}{\gamma^{2}}\,,
\end{equation}
where $\langle \hat{n} \rangle_{0}$ is the initial and $ \langle \hat{n} \rangle_{f}$ the final mean photon numbers in the cavity.
Thus $\Delta F$ of the cavity corresponds to the change in photon occupation of the cavity, as we would expect.

\begin{figure}[t]
    \centering
    \includegraphics[width=0.99\columnwidth]{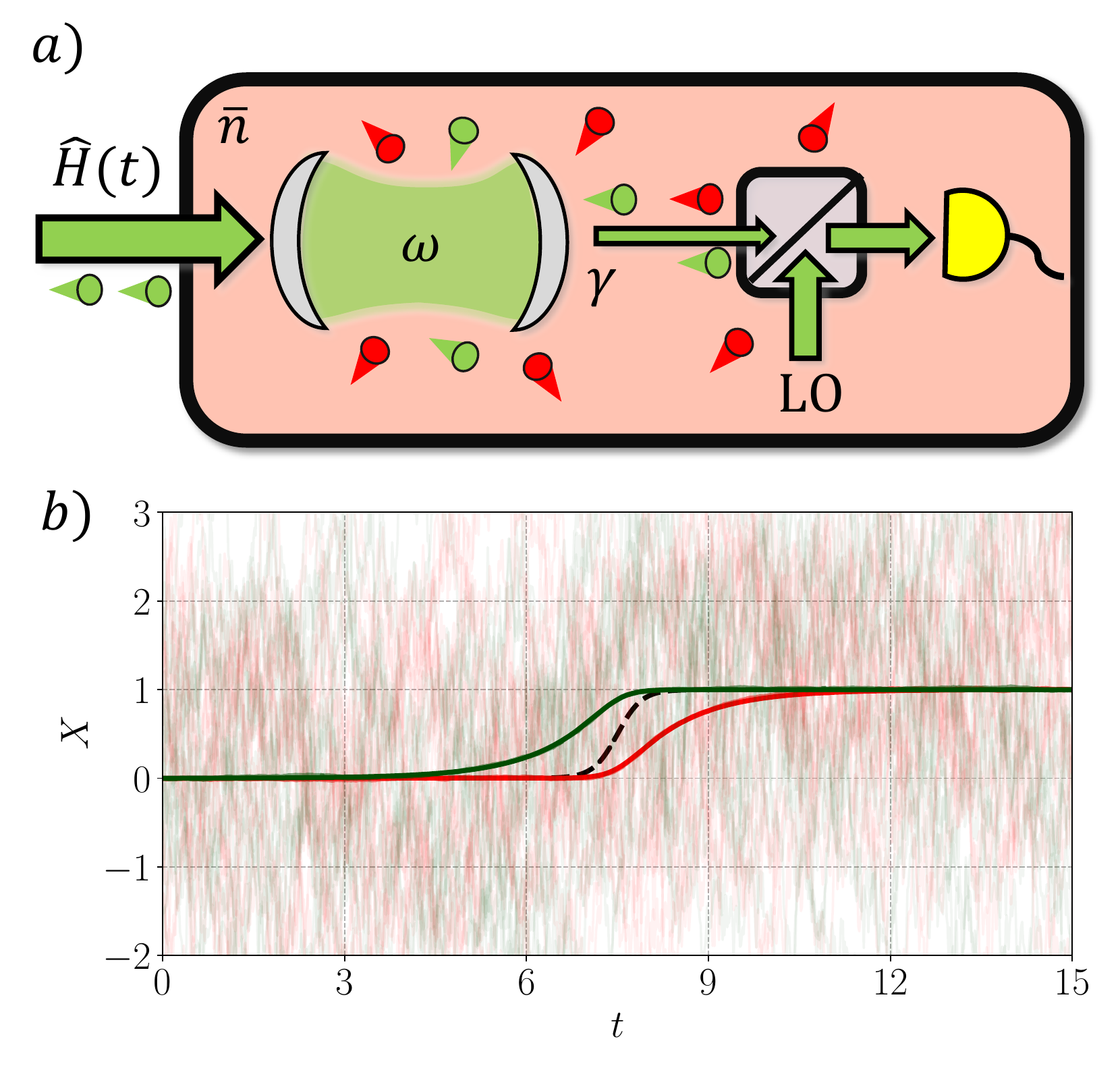}
    \caption{a) Here we consider a cavity with frequency $\omega$ that is coherently driven by a time-dependent field $\hat{H}(t)$ (green photons). The cavity is incoherently coupled to a thermal bath with mean photon number $\nbar$ and decay rate $\gamma$ (red photons). The thermal bath and the LO lead to white noise appearing at the measurement detector \cite{wiseman_quantum_2009}.
    b) The evolution of $X$ for the forward (red) and backward (green) simulation. The simulation parameters are the same as those used in Figure\,\ref{fig:NESS}. The solid corresponds to the average path taken corresponding to $\mathrm{E}[X] = \langle \hat{q}(t)\rangle$ and computed by the unconditional master equation. The opaque lines are a samples of estimate $X$ derived from the filtered homodyne current $J_{\mathrm{hom}}(t)$.
    Here we have used the same ramp $\lambda(t)$ used in Figure\,\ref{fig:NESS} and is depicted as the black dashed line.}
     \label{fig:figA}
\end{figure}

\section{Stochastic Quantum Thermodynamics and Measurements}
\label{sec:stochastic}
Now that we have described the dynamics and energetics of the system, we can turn to the measurement scheme.
In this example, we can continuously measure the work done on the system using homodyne detection of the phase quadrature $\hat{q}$ in Eq.\,(\ref{eq:work2}).
During the evolution of a quantum system, continuous quantum measurement conditions the evolution of the system $\hat{\rho}_{J}$ on the measurement outcome---where subscript $J$ indicates that this operator has been conditioned on the homodyne signal $J_{\mathrm{hom}}(t)$.
This stochastic conditioning is well understood for many continuous measurement protocols, including homodyne detection \cite{wiseman_quantum_2009}.
Here the evolution of conditional state $\hat{\rho}_{J}$ is described by an It\^o stochastic master equation with a white noise input \cite{wiseman_quantum_1993, wiseman_quantum_2009}
\begin{align}
    d \hat{\rho}_{J}&= - \frac{i}{\hbar} [\hat{H}_{I}, \hat{\rho}_{J}]dt \nonumber\\
    &+ \gamma(\nbar + 1)\mathcal{D}[\hat{a}]\hat{\rho}_{J}dt 
    +\gamma \nbar \mathcal{D}[\hat{a}^{\dagger}]\hat{\rho}_{J}dt \nonumber\\
    &+ \sqrt{\frac{\gamma \eta}{L}}\mathcal{H}[(\nbar+1)\hat{a} - \nbar \hat{a}^{\dagger}]\hat{\rho}_{J}d\mathcal{W}\,,  
\end{align}
where $d\mathcal{W}$ is a Wiener Process with $\mathrm{E}[d\mathcal{W}]{=} 0$, $\mathrm{Var}[d\mathcal{W}]{=}dt$, the coefficient $L {=} 2\overline{n}+1$, and the superoperator is $\mathcal{H}[\hat{a}]\hat{\rho} {= \hat{a} \hat{\rho} + \hat{\rho}\hat{a}^{\dagger} - \tr{\hat{a} \hat{\rho} + \hat{\rho}\hat{a}^{\dagger}}}$.
The homodyne measurement will provide a constant, albeit, stochastic readout of the phase quadrature $\langle \hat{q}_{J} \rangle$, which can be used to infer $\langle W \rangle$ via Eq.~(\ref{eq:work2}).
We can compute this analytically using the stochastic Heisenberg equation of motion for $\langle \hat{q}_{J} \rangle$ which is an It\^{o} stochastic differential equation (SDE)
\begin{align}
     d\langle \hat{q}_{J}\rangle &= \frac{g}{\omega}\lambda(t)dt - \frac{\gamma}{2}\langle \hat{q}_{J}\rangle dt\nonumber\\
     \label{eq:Heisenberg2}
     \quad & + \sqrt{\frac{2\gamma \eta\omega}{\hbar L}}\left(\Delta \hat{q}_{J}^{2} - \frac{\hbar L}{2\omega}\right) d\mathcal{W}\,,
\end{align}
where $\Delta \hat{q}_{J}^{2} = \langle \hat{q}^{2}_{J}\rangle - \langle \hat{q} _{J} \rangle^{2}$ is the conditional variance of $\hat{q}_{J}$ due to the measurement action on the quantum dynamics.
This setup is depicted fully in Figure\,\ref{fig:figA}a.

Given that our initial state is a thermal state, and thus a Gaussian state, the additional stochastic behaviour---which is Gaussian white noise---will preserve the Gaussian nature of our statistics in $\langle \hat{q}_{J}\rangle$.
As a result, the stochastic dynamics will only depend up to the second order moment $\Delta \hat{q}_{J}^{2}$ \cite{gardiner_stochastic_2009}.
Making use of It\^{o}'s lemma, we derive the Heisenberg equation of motion for the variance and we obtain
\begin{widetext}
\begin{align}
    d \Delta \hat{q}_{J}^{2} = -\gamma \Delta q_{J}^{2}dt 
    +\frac{\gamma \hbar L}{2\omega}dt -\frac{2\gamma \eta \omega }{\hbar L}\left(\Delta \hat{q}_{J}^{2} - \frac{\hbar L}{2\omega}\right)^{2}dt
    \nonumber 
    +
    \sqrt{\frac{2\gamma \eta\omega}{\hbar L}}d\mathcal{W}{\rm Skew}[\hat{q}_{J}]\,,
\end{align}
\end{widetext}
where ${\rm Skew}[\hat{q}_{J}] = \left(\langle \hat{q}_{J}^{3}\rangle  -  3\langle \hat{q}_{J}^{2}\rangle\langle \hat{q}_{J} \rangle + 2 \langle \hat{q} _{J} \rangle^{3}\right)$ is the third order moment (the skewness) which vanishes for Gaussian states.
Therefore the variance obeys an ordinary differential equation (ODE), which we can solve given the initial variance $\Delta \hat{q}_{J}^{2}(0) = \hbar L /2 \omega$. 
For all times, the solution to the variance ODE is
\begin{equation}
    \Delta \hat{q}_{J}^{2} =  \frac{\hbar L}{2 \omega}\,.
\end{equation}
If we now substitute this result into the SDE for $\langle \hat{q}_{J}\rangle$, the stochastic term also vanishes, permitting the general deterministic solution to the first moment Eq.\,(\ref{eq:Heisenberg2})
\begin{equation}
\label{eq:qj}
    \langle \hat{q}_{J}\rangle = \frac{g}{\omega}\int_{0}^{\tau}e^{-\gamma(\tau - t)/2}\lambda(t) dt\,.
\end{equation}
This tells us that the measurement apparatus does not condition the dynamics of the first moment. 
Using this result we can finally compute the average work done on the cavity by the oscillatory drive using Eq.\,(\ref{eq:work2})
\begin{equation}
\label{eq:avg_work}
    \langle W \rangle = g^{2}\int_{0}^{\tau}\int_{0}^{\tau}dt dt' \lambda(t)\lambda(t')e^{-\gamma(t-t')/2}\,.
\end{equation}

In a real experiment, the homodyne signal is generated by taking the continuum limit of the point process photocounts to a continuous photocurrent with white noise. If the local oscillator has classical intensity fluctuations then these will be cancelled out when the difference in the photocurrents is taken; this is commonly known as balanced homodyne detection.
Experimentally, we do not measure $\langle \hat{q}_{J}\rangle$ at each time step $t$, but rather infer a noisy classical estimate $X(t)$ which is derived from the observed homodyne current $J_{\mathrm{hom}}(t)$.
In an imperfect homodyne detection, we mix a strong local oscillator (LO) with the output of the cavity.
Once the LO has been subtracted out and removed, the signal homodyne current $J_\mathrm{hom}(t)$ \cite{wiseman_quantum_1993, wiseman_quantum_2009}
\begin{align}
\label{eq:current}
    J_{\mathrm{hom}}(t) &= \gamma \eta \langle \hat{a} + \hat{a}^{\dagger}\rangle_{J} + \sqrt{\gamma \eta L} \xi(t)\,,
\end{align}
where $0\leq \eta \leq 1$ is the measurement efficiency, $\xi(t) {=} d\mathcal{W}/dt$ is a delta correlated white noise and satisfies $\langle \xi(t)\rangle =0$ and $\langle \xi(t)\xi(t')\rangle{=}\delta(t-t')$ and has units of $1/\sqrt{dt}$.
Any fluctuations from the bath will show up as noise in our measurement result and limit the precision with which we can infer $\langle \hat{q}_{J}\rangle$.
Thus, there will necessarily be errors due to these fluctuations in our estimate of the average work $W$.

To determine the statistics of the work distributions, we must first find the distributions associated with our estimate $X(t)$.
This can be found by defining the associated SDE $dX(t)$ of the estimate obtained from the homodyne current $J_{\mathrm{hom}}(t)$, which is given by the classical It\^{o} SDE 
\begin{align}
\label{eq:SDE}
     d X(t) = \frac{g}{\omega}\lambda(t)dt - \frac{\gamma}{2}X(t) dt   + \sqrt{\frac{\gamma \hbar L}{2\omega \eta}} d\mathcal{W}\,,
\end{align}
where $d\mathcal{W}$ is another Wiener process. 
This SDE is a time-dependent Orstein-Ulenbeck process and preserves the Gaussian nature of $X(t)$ \cite{gardiner_stochastic_2009}.
The general solution to this SDE is 
\begin{align}
\label{eq:GeneralSDX}
    X(\tau) = X(0)e^{-\gamma \tau /2} + \langle \hat{q}_{J}\rangle +\sqrt{\frac{\gamma \hbar L}{2\omega \eta}}e^{-\gamma \tau/2} f(\tau)\,,
\end{align}
where $f(\tau) = \int_{0}^{\tau} d\mathcal{W} e^{\gamma t/2}$ is the integrated noise and has units of $\sqrt{dt}$.
We can therefore determine the probability distribution of each $X$ given $\langle \hat{q}_{J} \rangle$ at each time step using only the first two moments.
Unsurprisingly, the mean is trivial and satisfies $\overline{X} {=}\langle \hat{q}_{J} \rangle$ as we would expect given the initial condition $\overline{X}(0) {=} 0$.
In order to find the second order moment $\Delta X^{2}$ we need to look at the fluctuating term in Eq.\,(\ref{eq:GeneralSDX}), which for a non-anticipating function can be written as  \cite{gardiner_stochastic_2009}
\begin{equation}
    \Delta X^{2} = \frac{\gamma \hbar L}{2\omega \eta}e^{-\gamma \tau}f(\tau)^{2}\,.
\end{equation}
If we now take the average and make use the of the two-time correlation function of $f(t)$ \cite{gardiner_stochastic_2009}
\begin{equation}
\label{eq:two-time}
    \overline{f(t)f(t')} = \int_{0}^{\mathrm{min}(t,t')}d\tau e^{\gamma \tau}\,,
\end{equation}
then we obtain the average variance 
\begin{equation}
    \overline{\Delta X^{2}} = \frac{\gamma \hbar L}{2\omega \eta}\,.
\end{equation}
We can bring all this together to define the probability distribution over $X(t)$ given $\langle \hat{q}_{J} \rangle$ as the Gaussian distribution
\begin{equation}
\label{eq:PX}
    P(X(t)|\langle \hat{q}_{J}\rangle) = \mathcal{N}\exp\left(-\frac{\omega \eta (X(t) - \langle \hat{q}_{J}\rangle)^{2}}{ \hbar  L} \right)\,,
\end{equation}
where $\mathcal{N}=\sqrt{\omega \eta/\pi \hbar L}$ is the normalisation constant.
This distribution highlights a key component of quantum estimation theory which states that even when using a perfect measurement $\eta{=}1$ at zero temperature $L=1/2$, the precision of the estimate $X(t)$ of $\langle \hat{q}_{J}\rangle$ is bounded by the Heisenberg limit.
These estimated trajectories are numerically simulated and depicted in Figure \ref{fig:figA}b.

\begin{figure}[t]
    \centering
    \includegraphics[width=0.99\columnwidth]{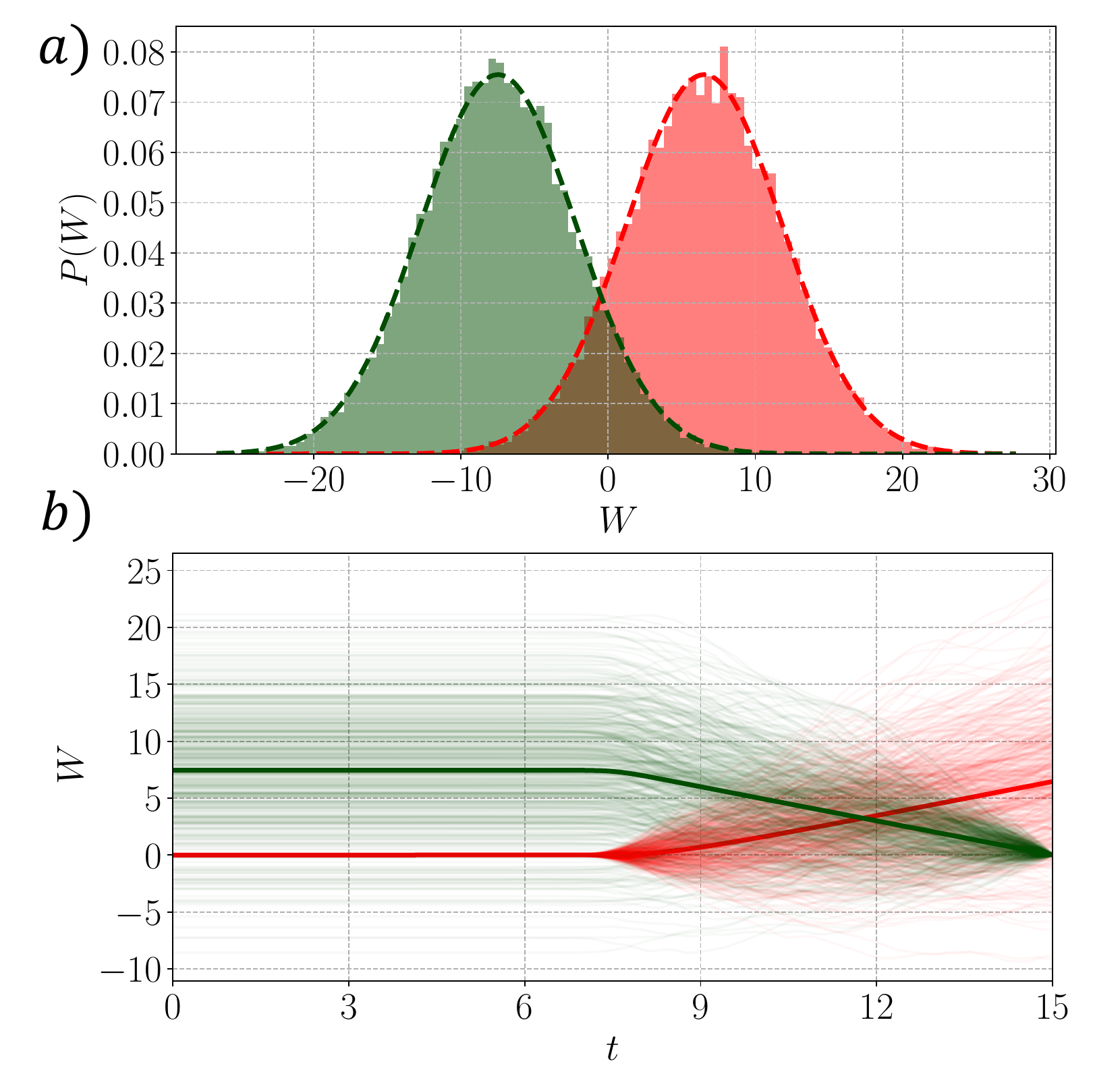}
    \caption{
    a) We plot the measured work distribution for both the forwards (red) and backwards (green) trajectories using a total of $20000$ simulations. Note they are not symmetric about $W=0$ which indicates a non-zero change in $\Delta F = -0.5$---for the simulation parameters---which corresponds to the intersection of these two distributions \cite{crooks_entropy_1999}.
    b) We plot $W$ as a function of time for the forwards and backwards trajectories. Given that $\hat{\rho}_{f}$ is a NESS, it requires constant work to maintain equilibrium. We can see that as time increases in the forward direction, the width in the estimated works diffuse due to the cumulative effects of energy fluctuations in the measurement signal, thus providing visual justification for inequality.\,(\ref{eq:Heisenberg_I}). In the reverse trajectories, the work at $\tau$ initially equals zero, but quickly diffuses as the drive is on. When the drive is turned off, the work stops at some finite value. }
    \label{fig:figB}
\end{figure}

We now turn to finding the estimated work distributions $P(W)$ using the general solution to the stochastic variable $X(t)$.
The \emph{estimated work} $W$ derived from the noisy signal $X(t)$ can be used to infer the average $\langle W \rangle$ determined by the quantum dynamics.
The estimated work $W$ is given in by integrating Eq.\,(\ref{eq:GeneralSDX}) 
\begin{equation}
\label{eq:workest}
    W = g\omega \int_{0}^{\tau} dt \lambda(t) X(t)\,.
\end{equation}
If we take the average of this estimate---and ignore the stochastic force term using the assumption $\overline{X}(0)=0$---we find that the inferred average work corresponds to the average work done on the cavity given by Eq.\,(\ref{eq:avg_work})
\begin{equation}
    \overline{W} = \langle W \rangle\,.
\end{equation}
Given that the work is also a non-anticipating function, the variance can also be defined by the stochastic term in Eq.\,(\ref{eq:workest}).
If we again make use of the two-time correlation function Eq.\,(\ref{eq:two-time}) and integrate, then we obtain the average variance in the estimate work 
\begin{align}
    \overline{\Delta W^{2}} &=  \frac{g^{2} \omega \hbar L}{ \eta}\int_{0}^{\tau}\int_{0}^{\tau}dt dt' \lambda(t)\lambda(t')e^{-\gamma(t-t')/2}\,,
\end{align}
which is more compactly written in terms of the average work
\begin{equation}
\label{eq:variance}
    \overline{\Delta W^{2}} = \frac{\hbar \omega L}{\eta}\langle W \rangle\,.
\end{equation}
If we now integrate over the range $t\in(0, \tau]$ such that the system has been given enough time to reach the NESS with $\lambda(\tau) {=}1$, then the inferred $W$ is well defined by a Gaussian distribution.
Thus, bringing all this together we can obtain an expression for the work distribution
\begin{equation}
\label{eq:work_dist}
    P(W) = \sqrt{\frac{\eta}{2\pi\hbar \omega L \langle W \rangle}}\exp\left(\frac{\eta (W - \langle W \rangle)^{2}}{2\hbar \omega L \langle W \rangle}\right)\,.
\end{equation}
The numerical simulation of these results along with the analytic prediction Eq.\,(\ref{eq:work_dist}) are depicted in Figure\,\ref{fig:figB}a showing perfect agreement. 

\section{Crooks' Theorem and Entropy Production}
We have so far described in detail the dynamics and estimation of the thermodynamic parameters needed to determine the $\Sigma$ in this model.
Now we can make use of Crooks' theorem \cite{crooks_entropy_1999} which tells us that the ratio of the forward $P_{F}(W)$ and the backward $P_{B}(-W)$ work distributions are directly related to the entropy production $\Sigma$
\begin{equation}
\label{eq:crooks}
    \frac{P_{F}(W)}{P_{B}(-W)} = e^{\Sigma}\,,
\end{equation}
which holds under all the assumptions outlined in Section\,\ref{sec:System Dynamics and Thermodynamics} and Section\,\ref{sec:stochastic}.
Here the reverse observations $\tilde{X}$ are generated by starting in the final state $\hat{\rho}_{f}$ with the drive on $\lambda(t){=}1$, and then reversing the ramp over the interval $t \in (\tau, 0]$, until the system settles back into the initial steady-state $\hat{\rho}_{i}$; this is otherwise known as a local time-reversal transformation \cite{seif_machine_2021}.
With the work distribution calculated in Eq.\,(\ref{eq:work_dist}), it is straight forward to compute the forward $P_{F}(W)$ and backward distribution $P_{R}(-W)$.
Taking the ratio of these distributions in accordance with Crook's theorem leads to 
\begin{equation}
\label{eq:pf}
    \frac{P_{F}(W)}{P_{B}(-W)} = \exp\left({\frac{\eta (W- \Delta F)}{\hbar \omega \left(\nbar +\frac{1}{2}\right)}}\right)\,,
\end{equation}
where $2\Delta F {=\langle W \rangle_{R} - \langle W \rangle_{F} }$ is the change in Helmholtz Free energy and coincides with the intersection of the two work distributions \cite{crooks_entropy_1999}.
Here $\langle W \rangle_{F}$ and $\langle W \rangle_{B}$ are the mean of the forward and backward distributions using Eq.\,(\ref{eq:work_dist}) respectively.
We have also assumed that $\Delta W_{F} \approx \Delta W_{R} $ which is valid in the limit where both the forwards and backwards dynamics have spent approximately equal time in their initial and final states as depicted in Figure\,\ref{fig:figB}a and Figure\,\ref{fig:figA}b.
Thus, we can directly associate the exponent with the entropy production 
\begin{equation}
    \Sigma = \frac{\eta (W - \Delta F)}{\hbar \omega\left(\nbar+\frac{1}{2}\right)}\,.
\end{equation}
We confirm these results by plotting the analytic solution derived above with the numerically fitted results using our simulations considered above. 
\begin{figure}[t]
    \centering
    \includegraphics[width=0.99\columnwidth]{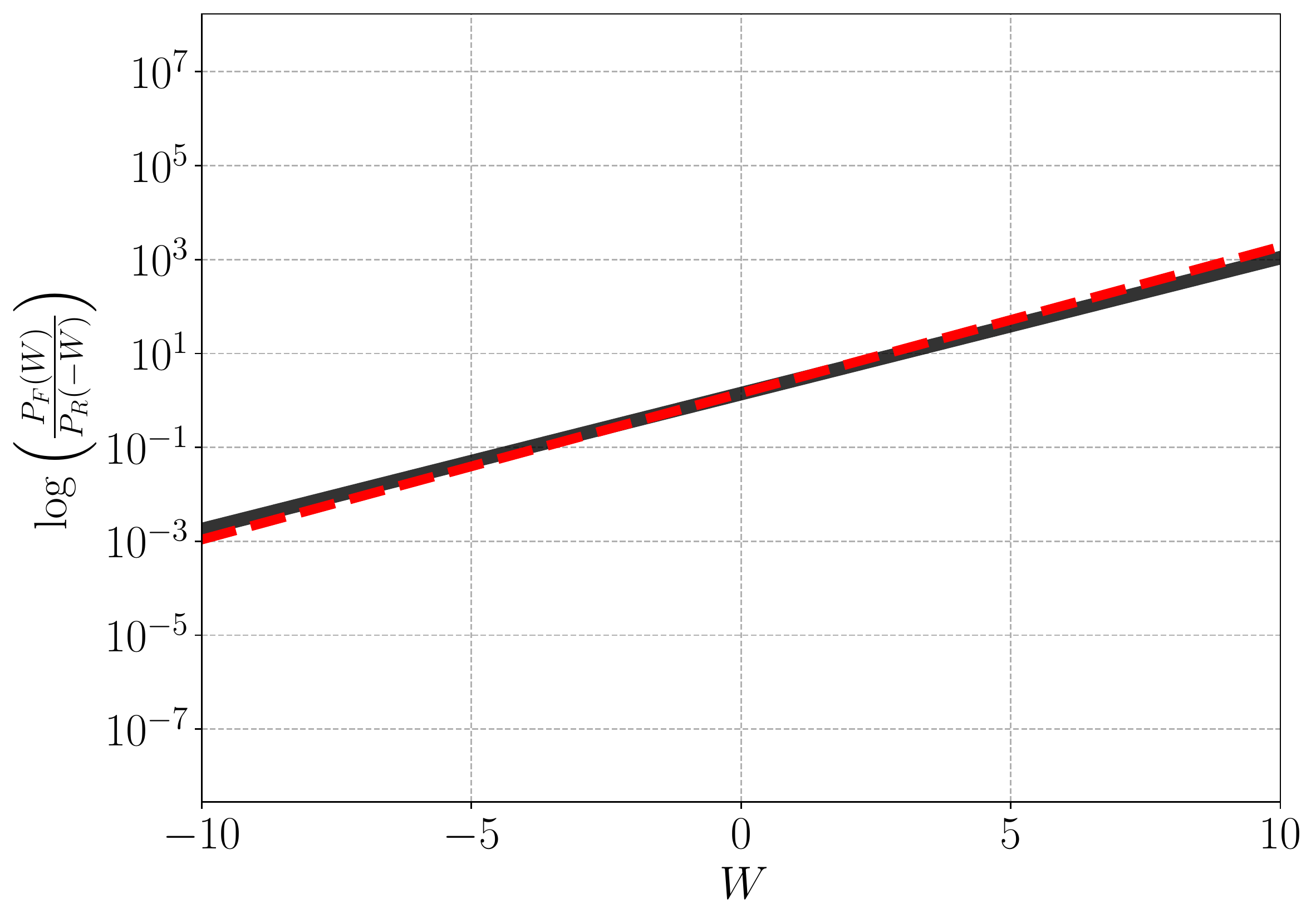}
    \caption{A log plot of the logarithm of the ratio of $P_{F}(W)/P_{B}(-W)$ as a function of the estimated $W$. The solid black line corresponds to our analytic results derived in Eq.\,(\ref{eq:pf}) where the red dashed line corresponds numerical fit derived from our simulations in Figure\,(\ref{fig:figA}). The gradient corresponds to $ \beta_{\mathrm{eff}}$ and the intercept corresponds to $- \beta_{\mathrm{eff}}\Delta F$. }
    \label{fig:crooks}
\end{figure}
The results of this are depicted in Figure\,\ref{fig:crooks} which shows the log plot of Eq.\,(\ref{eq:pf}) where the gradient corresponds to effective temperature
\begin{equation}
    \beta_{\mathrm{eff}} = \frac{\eta}{\hbar \omega \left( \nbar + \frac{1}{2}\right)}\,,
\end{equation}
which agrees---up to the measurement efficiency $\eta$---with that derived from the Wigner Entropy Ref.\,\cite{santos_wigner_2017} and ensures the entropy production does not diverge at zero temperature $\nbar{=}0$.
It is important to note that we are not arguing that $\beta_{\mathrm{eff}}$ is the inverse temperature---which is the bath temperature $T$---but that it is related to the stochastic fluctuations in the inferred work $W$ and is determined by the energy of the bath and the measurement efficiency. 
When $\nbar{=}0$, the presence of zero point energy ensures that $\beta_{\mathrm{eff}}{=}2/\hbar\omega$ and thus leads to a well defined entropy production.

Unlike the Wigner Entropy derivation, our derivation shows that $\beta_{\mathrm{eff}}$ depends explicitly on the measurement efficiency, highlighting that the effective temperature is directly related to the measurement record.
When $\eta=0$, no information about the work is gained, and thus the measured entropy production in zero. 
Given that the effective inverse temperature $\beta_{\mathrm{eff}}$ is determined by the fluctuations in the measurement record, we can therefore think of this as the entropy produced to acquire the estimate of the work. 

We can now compute the average entropy production in the forwards direction by computing the relative Shannon entropy of the work distributions from our Eq.\,(\ref{eq:pf}) 
\begin{align}
\label{eq:Second_law}
    \overline{\Sigma} &= \int dW P_{F}(W)\log \left(\frac{P_{F}(W)}{P_{B}(-W)}\right) \,,\\
\label{eq:EP}
    &= \beta_{\mathrm{eff}}\left(\overline{W}_{F} - \Delta F\right)\,.
\end{align}
We further verify this result by numerically finding the ratio of the forward and backward work distributions plotted in Figure\,\ref{fig:figB}a and plotting it against the Eq.\,(\ref{eq:EP}) which is depicted in Figure\,\ref{fig:crooks}. 
Moreover it allows us to define the inequality of the entropy production
\begin{equation}
\label{eq:TUR}
    \overline{\Sigma} \geq  \beta_{\mathrm{eff}} \overline{W}_{F}\,.
\end{equation}
On closer inspection, one will recognise the RHS of this inequality as equal to $\overline{\Sigma} {\geq} 2\langle W \rangle_{F}^{2}/\overline{\Delta W_{F}^{2}}$, where the variance is given by Eq.\,(\ref{eq:variance}), which is consistent with TURs \cite{horowitz_thermodynamic_2020}.
Thermodynamic uncertainty relations quantify the fundamental lower bound on the trade-off between precision with which thermodynamic quantities can be measured---namely the signal to noise ratio (SNR)---and entropy production. 
This precision is related to the integrated current exchanged during an out-of-equilibrium process over some time interval, which in this case is related to the average work inferred via the measured homodyne signal.
Therefore we can think of Eq.\,(\ref{eq:Second_law}) as a TUR that is the minimum amount of entropy produced when estimating the work $ \overline{W}_{F}$ in a temperature quantum system under the described dynamics.
Furthermore, as the average amount of work increases, so to does the entropy which corresponds to a decrease in the SNR. 
This is due to the diffusive nature of work distribution Eq.\,(\ref{eq:work_dist}) and can be seen clearly in the Figure\,\ref{fig:figB}c. 
When driven far from equilibrium the cavity requires constant work be done to maintain the NESS. 
However due to the measurement fluctuations arising from the homodyne signal, this estimate is stochastic and diffuses in time.

\section{Classical Fisher Information (FI) and a new inequality}
We have thus far characterised thermodynamic quantities such as work and entropy production in a low temperature, quantum optical cavity, by applying classical estimation theory to the homodyne detection measurement record. 
We now quantify how much signal (or information) is carried by our estimation $W$ of the average work done $\langle W \rangle_{F}$ in the forward process, using classical FI $I(\langle W \rangle_{F})$.
Mathematically, the FI of the work distribution is defined
\begin{equation}
     I(\overline{W}_{F}) = -\mathrm{E}\left[\left.\frac{\partial^{2}}{\partial  \overline{W}^{2}} \log (P(W))\right|\overline{W}_{F}\right]\,,
\end{equation}
which we can compute using Eq.\,(\ref{eq:work_dist}) to obtain
\begin{equation}
\label{eq:Heisenberg_I}
    I(\overline{W}_{F}) = \frac{ \beta_{\mathrm{eff}}}{2\overline{W}_{F}} + \frac{1}{\overline{W}_{F}^{2}} \geq \frac{1}{\overline{\Delta W^{2}}}\,,
\end{equation}
which is the Cram\'er-Rao inequality for large $\overline{W}_{F}$  as the estimated work becomes an unbiased estimator. 
The Cram\'er-Rao inequality tells us that the precision with which we can estimate the average work $\overline{W}_{F}$ decreases with the amount of work done due to cumulative effect of fluctuations in our measurement record. 
As an interesting aside, this result may be useful in understanding the relation to the precision in time keeping, which is the conjugate variable of energy. 
As more heat is dissipated, the Fisher information in the energy estimate decreases, but arguably increases in the precision of time \cite{erker_autonomous_2017, milburn_thermodynamics_2020, pearson_measuring_2021}.
The effect of this can be seen in the diffusive nature of the estimated work in Figure\,\ref{fig:figA}d for increasing time. 

\begin{figure}
    \centering
    \includegraphics[width=\columnwidth]{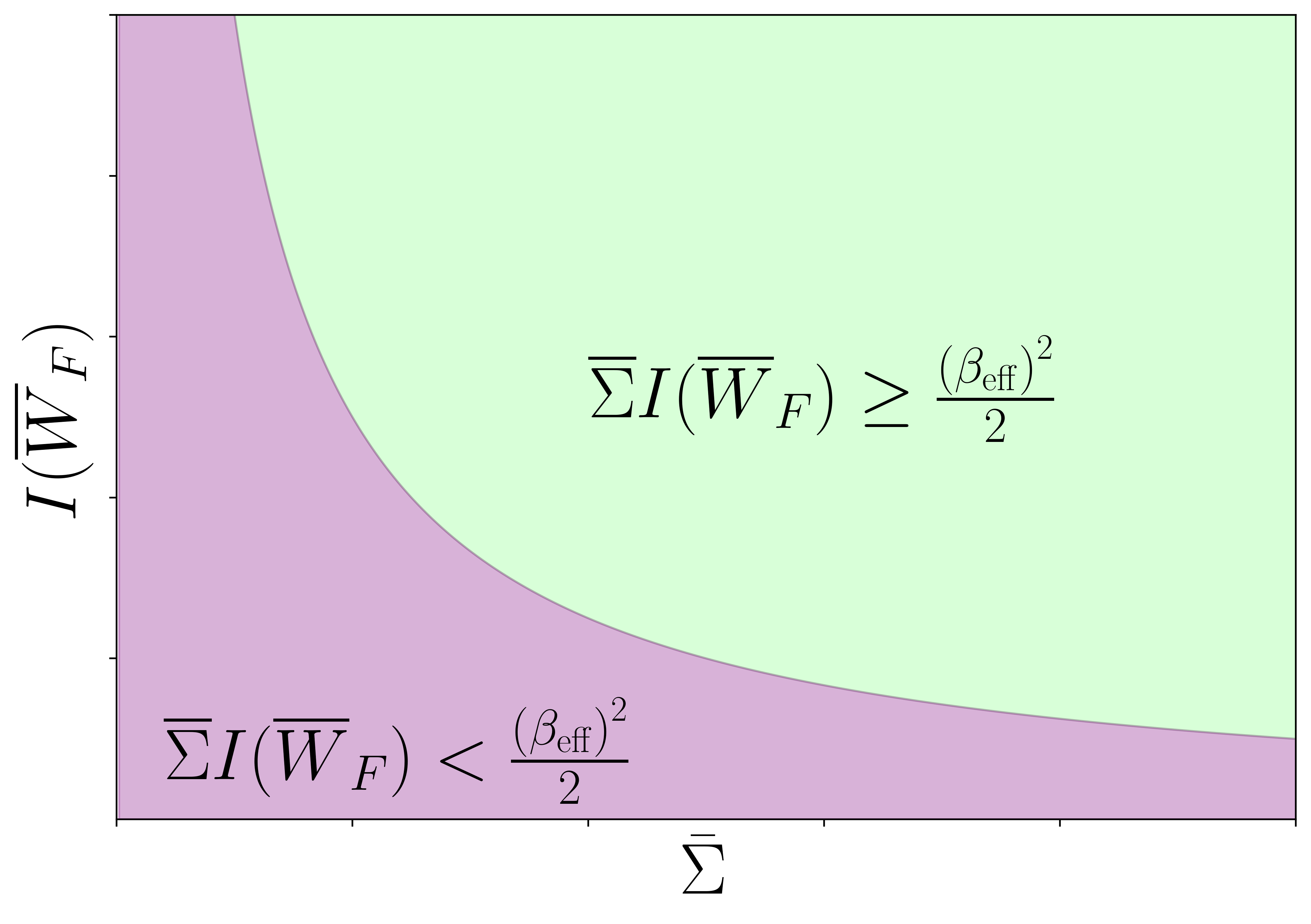}
    \caption{Trade off between classical FI and the entropy production of measurements of the average work $\overline{W}$. The green region is satisfies the inequality, whereas the purple region does not. }
    \label{fig:inequality}
\end{figure}

By itself, the Cram\'er-Rao inequality alone is unsurprising, but we can now go a step further and combine the quantum inequality\,(\ref{eq:Second_law}) and the Cram\'er-Rao inequality \,(\ref{eq:Heisenberg_I}) under a single bound
\begin{equation}
\label{eq:information}
    I(\overline{W}_{F})\overline{\Sigma} \geq \frac{ \beta_{\mathrm{eff}}^{2}}{2}\,.
\end{equation}
This inequality tells us that the product of the information about $\overline{W}_{F}$ which is obtained by producing entropy $\overline{\Sigma}$ must always be greater than half of the squared effective inverse temperature.
The interpretation of this inequality is straight forward: estimating a thermodynamic variable such as work $\overline{W}_{F}$ requires that a minimum amount of entropy be produced in the measurement process.
However this measurement is not error free due to cumulative effect of fluctuations in the measurement record due to non-zero effective temperature of the bath.
Thus, greater information $I(\overline{W}_{F})$ requires lowering the average entropy production, but at the quantum limit with unit efficiency $\eta{=}1$, this cannot be reduced to zero: there will always be errors in the measurement record producing entropy as depicted in Figure\,\ref{fig:inequality}.

\section{Conclusion}

In this article we have sought to establish how the entropy production for zero temperature quantum systems can be defined.
Using stochastic quantum thermodynamics we have shown that for a zero temperature optical cavity subject to homodyne detection, the inverse temperature $\beta$ is replaced by the effective inverse temperature $\beta_{\mathrm{eff}}$ which is related to the zero point energy of the environment. This effective temperature does not diverge at zero temperature and thus ensures that the estimated entropy production is well defined even as the environment approaches zero temperature. This result further corroborates the effective temperature derived using the Wigner entropy \cite{santos_wigner_2017, belenchia_entropy_2020}. The implication of this work is that it is physically meaningful to define entropy production using continuous measurement records and estimations of thermodynamic quantities. In doing so, the role of the measurement device is seen to be critical, as it sets the nature of the measured fluctuations. 
Furthermore, the entropy contained in the measurement record connects Crooks' fluctuation theorem at the quantum limit to the Cram\'er Rao bound, under a single information inequality (\ref{eq:information}). This inequality fundamentally bounds the trade-off between Fisher information and entropy production for a zero temperature quantum system subject to continuous homodyne measurement. From this we can see that measuring a higher average entropy production necessarily implies lower FI in the measured signal. Thus higher entropy coincides with lower precision and agrees with the standard interpretation of the TUR. 
Finally, our results hold for Gaussian states, weakly coupled to the bath, with a linear Hamiltonian. It will be interesting to consider in future work whether it also holds in the strong coupling limit \cite{miller_energy-temperature_2018}, and for higher order Hamiltonians containing dissipative phase transitions \cite{goes_entropy_2020, goes_quantum_2020}.

\section{Acknowledgements}
We would like to thank Gerard Milburn for many illuminating discussions and recommendations. This work was supported by the Australian Research Council Centre of Excellence for Engineered Quantum Systems (Project No.CE170100009), and the Foundational Questions Institute Fund (Project No. FQXi-IAF19-04). 

\bibliographystyle{quantum}
\bibliography{Thermodynamics}

\begin{thebibliography}{10}

\bibitem{goold_role_2016}
John Goold, Marcus Huber, Arnau Riera, Lídia~del Rio, and Paul Skrzypczyk.
\newblock ``The role of quantum information in thermodynamics—a topical
  review''.
\newblock \href{https://dx.doi.org/10.1088/1751-8113/49/14/143001}{Journal of
  Physics A: Mathematical and Theoretical {\bf 49}, 143001}~(2016).

\bibitem{vinjanampathy_quantum_2016}
Sai Vinjanampathy and Janet Anders.
\newblock ``Quantum thermodynamics''.
\newblock \href{https://dx.doi.org/10.1080/00107514.2016.1201896}{Contemporary
  Physics {\bf 57}, 545--579}~(2016).

\bibitem{seifert_stochastic_2012}
Udo Seifert.
\newblock ``Stochastic thermodynamics, fluctuation theorems and molecular
  machines''.
\newblock \href{https://dx.doi.org/10.1088/0034-4885/75/12/126001}{Reports on
  Progress in Physics {\bf 75}, 126001}~(2012).

\bibitem{jarzynski_nonequilibrium_1997}
C.~Jarzynski.
\newblock ``Nonequilibrium {Equality} for {Free} {Energy} {Differences}''.
\newblock \href{https://dx.doi.org/10.1103/PhysRevLett.78.2690}{Physical Review
  Letters {\bf 78}, 2690--2693}~(1997).

\bibitem{crooks_entropy_1999}
Gavin~E. Crooks.
\newblock ``Entropy production fluctuation theorem and the nonequilibrium work
  relation for free energy differences''.
\newblock \href{https://dx.doi.org/10.1103/PhysRevE.60.2721}{Physical Review E
  {\bf 60}, 2721--2726}~(1999).

\bibitem{hatano_steady-state_2001}
Takahiro Hatano and Shin-ichi Sasa.
\newblock ``Steady-{State} {Thermodynamics} of {Langevin} {Systems}''.
\newblock \href{https://dx.doi.org/10.1103/PhysRevLett.86.3463}{Physical Review
  Letters {\bf 86}, 3463--3466}~(2001).

\bibitem{seifert_entropy_2005}
Udo Seifert.
\newblock ``Entropy {Production} along a {Stochastic} {Trajectory} and an
  {Integral} {Fluctuation} {Theorem}''.
\newblock \href{https://dx.doi.org/10.1103/PhysRevLett.95.040602}{Physical
  Review Letters {\bf 95}, 040602}~(2005).

\bibitem{horowitz_thermodynamic_2020}
Jordan~M. Horowitz and Todd~R. Gingrich.
\newblock ``Thermodynamic uncertainty relations constrain non-equilibrium
  fluctuations''.
\newblock \href{https://dx.doi.org/10.1038/s41567-019-0702-6}{Nature Physics
  {\bf 16}, 15--20}~(2020).

\bibitem{strasberg_operational_2019}
Philipp Strasberg.
\newblock ``Operational approach to quantum stochastic thermodynamics''.
\newblock \href{https://dx.doi.org/10.1103/PhysRevE.100.022127}{Physical Review
  E {\bf 100}, 022127}~(2019).

\bibitem{strasberg_non-markovianity_2019}
Philipp Strasberg and Massimiliano Esposito.
\newblock ``Non-{Markovianity} and negative entropy production rates''.
\newblock \href{https://dx.doi.org/10.1103/PhysRevE.99.012120}{Physical Review
  E {\bf 99}, 012120}~(2019).

\bibitem{davies_quantum_1969}
E.~B. Davies.
\newblock ``Quantum stochastic processes''.
\newblock \href{https://dx.doi.org/10.1007/BF01645529}{Communications in
  Mathematical Physics {\bf 15}, 277--304}~(1969).

\bibitem{gardiner_quantum_2004}
Crispin Gardiner and Peter Zoller.
\newblock ``Quantum noise: A handbook of markovian and non-markovian quantum
  stochastic methods with applications to quantum optics''.
\newblock Springer Science and Business Media. ~(2004).

\bibitem{wiseman_quantum_2009}
Howard~M. Wiseman and Gerard~J. Milburn.
\newblock ``Quantum {Measurement} and {Control}''.
\newblock \href{https://dx.doi.org/10.1017/CBO9780511813948}{Cambridge
  University Press}. ~(2009).

\bibitem{mukamel_quantum_2003}
Shaul Mukamel.
\newblock ``Quantum {Extension} of the {Jarzynski} {Relation}: {Analogy} with
  {Stochastic} {Dephasing}''.
\newblock \href{https://dx.doi.org/10.1103/PhysRevLett.90.170604}{Physical
  Review Letters {\bf 90}, 170604}~(2003).

\bibitem{popescu_entanglement_2006}
Sandu Popescu, Anthony~J. Short, and Andreas Winter.
\newblock ``Entanglement and the foundations of statistical mechanics''.
\newblock \href{https://dx.doi.org/10.1038/nphys444}{Nature Physics {\bf 2},
  754--758}~(2006).

\bibitem{esposito_nonequilibrium_2009}
Massimiliano Esposito, Upendra Harbola, and Shaul Mukamel.
\newblock ``Nonequilibrium fluctuations, fluctuation theorems, and counting
  statistics in quantum systems''.
\newblock \href{https://dx.doi.org/10.1103/RevModPhys.81.1665}{Reviews of
  Modern Physics {\bf 81}, 1665--1702}~(2009).

\bibitem{landi_irreversible_2020}
Gabriel~T. Landi and Mauro Paternostro.
\newblock ``Irreversible entropy production: From classical to quantum''.
\newblock \href{https://dx.doi.org/10.1103/RevModPhys.93.035008}{Rev. Mod.
  Phys. {\bf 93}, 035008}~(2021).

\bibitem{campisi_colloquium_2011}
Michele Campisi, Peter Hänggi, and Peter Talkner.
\newblock ``Colloquium: {Quantum} fluctuation relations: {Foundations} and
  applications''.
\newblock \href{https://dx.doi.org/10.1103/RevModPhys.83.771}{Reviews of Modern
  Physics {\bf 83}, 771--791}~(2011).

\bibitem{hasegawa_thermodynamic_2021}
Yoshihiko Hasegawa.
\newblock ``Thermodynamic {Uncertainty} {Relation} for {General} {Open}
  {Quantum} {Systems}''.
\newblock \href{https://dx.doi.org/10.1103/PhysRevLett.126.010602}{Physical
  Review Letters {\bf 126}, 010602}~(2021).

\bibitem{miller_joint_2021}
Harry J.~D. Miller, M.~Hamed Mohammady, Mart\'{\i} Perarnau-Llobet, and Giacomo
  Guarnieri.
\newblock ``Joint statistics of work and entropy production along quantum
  trajectories''.
\newblock \href{https://dx.doi.org/10.1103/PhysRevE.103.052138}{Phys. Rev. E
  {\bf 103}, 052138}~(2021).

\bibitem{Stefano_noequilibrium_2018}
P.~G. Di~Stefano, J.~J. Alonso, E.~Lutz, G.~Falci, and M.~Paternostro.
\newblock ``Nonequilibrium thermodynamics of continuously measured quantum
  systems: A circuit qed implementation''.
\newblock \href{https://dx.doi.org/10.1103/PhysRevB.98.144514}{Phys. Rev. B
  {\bf 98}, 144514}~(2018).

\bibitem{horowitz_quantum-trajectory_2012}
Jordan~M. Horowitz.
\newblock ``Quantum-trajectory approach to the stochastic thermodynamics of a
  forced harmonic oscillator''.
\newblock \href{https://dx.doi.org/10.1103/PhysRevE.85.031110}{Physical Review
  E {\bf 85}, 031110}~(2012).

\bibitem{horowitz_entropy_2013}
Jordan~M. Horowitz and Juan M.~R. Parrondo.
\newblock ``Entropy production along nonequilibrium quantum jump
  trajectories''.
\newblock \href{https://dx.doi.org/10.1088/1367-2630/15/8/085028}{New Journal
  of Physics {\bf 15}, 085028}~(2013).

\bibitem{alonso_thermodynamics_2016}
Jose~Joaquin Alonso, Eric Lutz, and Alessandro Romito.
\newblock ``Thermodynamics of {Weakly} {Measured} {Quantum} {Systems}''.
\newblock \href{https://dx.doi.org/10.1103/PhysRevLett.116.080403}{Physical
  Review Letters {\bf 116}, 080403}~(2016).

\bibitem{elouard_role_2017}
Cyril Elouard, David~A. Herrera-Martí, Maxime Clusel, and Alexia Auffèves.
\newblock ``The role of quantum measurement in stochastic thermodynamics''.
\newblock \href{https://dx.doi.org/10.1038/s41534-017-0008-4}{npj Quantum
  Information{\bf 3}}~(2017).

\bibitem{manikandan_fluctuation_2019}
Sreenath~K. Manikandan, Cyril Elouard, and Andrew~N. Jordan.
\newblock ``Fluctuation theorems for continuous quantum measurements and
  absolute irreversibility''.
\newblock \href{https://dx.doi.org/10.1103/PhysRevA.99.022117}{Phys. Rev. A
  {\bf 99}, 022117}~(2019).

\bibitem{belenchia_entropy_2020}
Alessio Belenchia, Luca Mancino, Gabriel~T. Landi, and Mauro Paternostro.
\newblock ``Entropy production in continuously measured {Gaussian} quantum
  systems''.
\newblock \href{https://dx.doi.org/10.1038/s41534-020-00334-6}{npj Quantum
  Information{\bf 6}}~(2020).

\bibitem{naghiloo_heat_2020}
M.~Naghiloo, D.~Tan, P.~M. Harrington, J.~J. Alonso, E.~Lutz, A.~Romito, and
  K.~W. Murch.
\newblock ``Heat and {Work} {Along} {Individual} {Trajectories} of a {Quantum}
  {Bit}''.
\newblock \href{https://dx.doi.org/10.1103/PhysRevLett.124.110604}{Physical
  Review Letters {\bf 124}, 110604}~(2020).

\bibitem{guryanova_ideal_2020}
Yelena Guryanova, Nicolai Friis, and Marcus Huber.
\newblock ``Ideal {Projective} {Measurements} {Have} {Infinite} {Resource}
  {Costs}''.
\newblock \href{https://dx.doi.org/10.22331/q-2020-01-13-222}{Quantum {\bf 4},
  222}~(2020).

\bibitem{elouard_extracting_2017}
Cyril Elouard, David Herrera-Martí, Benjamin Huard, and Alexia Auffèves.
\newblock ``Extracting {Work} from {Quantum} {Measurement} in {Maxwell}’s
  {Demon} {Engines}''.
\newblock \href{https://dx.doi.org/10.1103/PhysRevLett.118.260603}{Physical
  Review Letters {\bf 118}, 260603}~(2017).

\bibitem{naghiloo_information_2018}
M.~Naghiloo, J.~J. Alonso, A.~Romito, E.~Lutz, and K.~W. Murch.
\newblock ``Information {Gain} and {Loss} for a {Quantum} {Maxwell}'s
  {Demon}''.
\newblock \href{https://dx.doi.org/10.1103/PhysRevLett.121.030604}{Physical
  Review Letters {\bf 121}, 030604}~(2018).

\bibitem{monsel_energetic_2020}
Juliette Monsel, Marco Fellous-Asiani, Benjamin Huard, and Alexia Auffèves.
\newblock ``The {Energetic} {Cost} of {Work} {Extraction}''.
\newblock \href{https://dx.doi.org/10.1103/PhysRevLett.124.130601}{Physical
  Review Letters {\bf 124}, 130601}~(2020).

\bibitem{mitchison_charging_2021}
Mark~T. Mitchison, John Goold, and Javier Prior.
\newblock ``Charging a quantum battery with linear feedback control''.
\newblock \href{https://dx.doi.org/10.22331/q-2021-07-13-500}{Quantum{\bf
  5}}~(2021).

\bibitem{hasegawa_quantum_2020}
Yoshihiko Hasegawa.
\newblock ``Quantum {Thermodynamic} {Uncertainty} {Relation} for {Continuous}
  {Measurement}''.
\newblock \href{https://dx.doi.org/10.1103/PhysRevLett.125.050601}{Physical
  Review Letters {\bf 125}, 050601}~(2020).

\bibitem{barato_thermodynamic_2015}
Andre~C. Barato and Udo Seifert.
\newblock ``Thermodynamic {Uncertainty} {Relation} for {Biomolecular}
  {Processes}''.
\newblock \href{https://dx.doi.org/10.1103/PhysRevLett.114.158101}{Physical
  Review Letters {\bf 114}, 158101}~(2015).

\bibitem{pietzonka_universal_2016}
Patrick Pietzonka, Andre~C. Barato, and Udo Seifert.
\newblock ``Universal bounds on current fluctuations''.
\newblock \href{https://dx.doi.org/10.1103/PhysRevE.93.052145}{Physical Review
  E {\bf 93}, 052145}~(2016).

\bibitem{pietzonka_universal_2018}
Patrick Pietzonka and Udo Seifert.
\newblock ``Universal {Trade}-{Off} between {Power}, {Efficiency}, and
  {Constancy} in {Steady}-{State} {Heat} {Engines}''.
\newblock \href{https://dx.doi.org/10.1103/PhysRevLett.120.190602}{Physical
  Review Letters {\bf 120}, 190602}~(2018).

\bibitem{van_vu_thermodynamic_2020}
Tan Van~Vu and Yoshihiko Hasegawa.
\newblock ``Thermodynamic uncertainty relations under arbitrary control
  protocols''.
\newblock \href{https://dx.doi.org/10.1103/PhysRevResearch.2.013060}{Physical
  Review Research {\bf 2}, 013060}~(2020).

\bibitem{vu_uncertainty_2020}
Tan~Van Vu and Yoshihiko Hasegawa.
\newblock ``Uncertainty relation under information measurement and feedback
  control''.
\newblock \href{https://dx.doi.org/10.1088/1751-8121/ab64a4}{Journal of Physics
  A: Mathematical and Theoretical {\bf 53}, 075001}~(2020).

\bibitem{potts_thermodynamic_2019}
Patrick~P. Potts and Peter Samuelsson.
\newblock ``Thermodynamic uncertainty relations including measurement and
  feedback''.
\newblock \href{https://dx.doi.org/10.1103/PhysRevE.100.052137}{Physical Review
  E {\bf 100}, 052137}~(2019).

\bibitem{debiossac_thermodynamics_2020}
Maxime Debiossac, David Grass, Jose~Joaquin Alonso, Eric Lutz, and Nikolai
  Kiesel.
\newblock ``Thermodynamics of continuous non-{Markovian} feedback control''.
\newblock \href{https://dx.doi.org/10.1038/s41467-020-15148-5}{Nature
  Communications{\bf 11}}~(2020).

\bibitem{cao_thermodynamics_2009}
F.~J. Cao and M.~Feito.
\newblock ``Thermodynamics of feedback controlled systems''.
\newblock \href{https://dx.doi.org/10.1103/PhysRevE.79.041118}{Physical Review
  E {\bf 79}, 041118}~(2009).

\bibitem{horowitz_nonequilibrium_2010}
Jordan~M. Horowitz and Suriyanarayanan Vaikuntanathan.
\newblock ``Nonequilibrium detailed fluctuation theorem for repeated discrete
  feedback''.
\newblock \href{https://dx.doi.org/10.1103/PhysRevE.82.061120}{Physical Review
  E {\bf 82}, 061120}~(2010).

\bibitem{Gong_quantum_2016}
Zongping Gong, Yuto Ashida, and Masahito Ueda.
\newblock ``Quantum-trajectory thermodynamics with discrete feedback control''.
\newblock \href{https://dx.doi.org/10.1103/PhysRevA.94.012107}{Phys. Rev. A
  {\bf 94}, 012107}~(2016).

\bibitem{murashita_fluctuation_2017}
Y\^uto Murashita, Zongping Gong, Yuto Ashida, and Masahito Ueda.
\newblock ``Fluctuation theorems in feedback-controlled open quantum systems:
  Quantum coherence and absolute irreversibility''.
\newblock \href{https://dx.doi.org/10.1103/PhysRevA.96.043840}{Phys. Rev. A
  {\bf 96}, 043840}~(2017).

\bibitem{manzano_quantum_2021}
Gonzalo Manzano and Roberta Zambrini.
\newblock ``Quantum thermodynamics under continuous monitoring: a general
  framework''~(2021).
\newblock  \href{http://arxiv.org/abs/2112.02019}{arXiv:2112.02019}.

\bibitem{masanes_general_2017}
Lluís Masanes and Jonathan Oppenheim.
\newblock ``A general derivation and quantification of the third law of
  thermodynamics''.
\newblock \href{https://dx.doi.org/10.1038/ncomms14538}{Nature
  Communications{\bf 8}}~(2017).

\bibitem{santos_wigner_2017}
Jader~P. Santos, Gabriel~T. Landi, and Mauro Paternostro.
\newblock ``Wigner {Entropy} {Production} {Rate}''.
\newblock \href{https://dx.doi.org/10.1103/PhysRevLett.118.220601}{Physical
  Review Letters {\bf 118}, 220601}~(2017).

\bibitem{adesso_measuring_2012}
Gerardo Adesso, Davide Girolami, and Alessio Serafini.
\newblock ``Measuring {Gaussian} {Quantum} {Information} and {Correlations}
  {Using} the {Rényi} {Entropy} of {Order} 2''.
\newblock \href{https://dx.doi.org/10.1103/PhysRevLett.109.190502}{Physical
  Review Letters {\bf 109}, 190502}~(2012).

\bibitem{wiseman_quantum_1993}
H.~M. Wiseman and G.~J. Milburn.
\newblock ``Quantum theory of field-quadrature measurements''.
\newblock \href{https://dx.doi.org/10.1103/PhysRevA.47.642}{Physical Review A
  {\bf 47}, 642--662}~(1993).

\bibitem{gardiner_stochastic_2009}
Crispin Gardiner.
\newblock ``Stochastic {Methods}: {A} {Handbook} for the {Natural} and {Social}
  {Sciences}''.
\newblock Springer {Series} in {Synergetics}. Springer-Verlag. Berlin
  Heidelberg~(2009).
\newblock 4 edition.
\newblock
  url:~\href{https://www.springer.com/gp/book/9783540707127}{www.springer.com/gp/book/9783540707127}.

\bibitem{seif_machine_2021}
Alireza Seif, Mohammad Hafezi, and Christopher Jarzynski.
\newblock ``Machine learning the thermodynamic arrow of time''.
\newblock \href{https://dx.doi.org/10.1038/s41567-020-1018-2}{Nature
  Physics{\bf 17}}~(2021).

\bibitem{erker_autonomous_2017}
Paul Erker, Mark~T. Mitchison, Ralph Silva, Mischa~P. Woods, Nicolas Brunner,
  and Marcus Huber.
\newblock ``Autonomous {Quantum} {Clocks}: {Does} {Thermodynamics} {Limit}
  {Our} {Ability} to {Measure} {Time}?''.
\newblock \href{https://dx.doi.org/10.1103/PhysRevX.7.031022}{Physical Review X
  {\bf 7}, 031022}~(2017).

\bibitem{milburn_thermodynamics_2020}
G.~J. Milburn.
\newblock ``The thermodynamics of clocks''.
\newblock \href{https://dx.doi.org/10.1080/00107514.2020.1837471}{Contemporary
  Physics{\bf 61}}~(2020).

\bibitem{pearson_measuring_2021}
A.~N. Pearson, Y.~Guryanova, P.~Erker, E.~A. Laird, G.~A.~D. Briggs, M.~Huber,
  and N.~Ares.
\newblock ``Measuring the {Thermodynamic} {Cost} of {Timekeeping}''.
\newblock \href{https://dx.doi.org/10.1103/PhysRevX.11.021029}{Physical Review
  X {\bf 11}, 021029}~(2021).

\bibitem{miller_energy-temperature_2018}
H.~J.~D. Miller and J.~Anders.
\newblock ``Energy-temperature uncertainty relation in quantum
  thermodynamics''.
\newblock \href{https://dx.doi.org/10.1038/s41467-018-04536-7}{Nature
  Communications {\bf 9}, 2203}~(2018).

\bibitem{goes_entropy_2020}
Bruno~O. Goes and Gabriel~T. Landi.
\newblock ``Entropy production dynamics in quench protocols of a
  driven-dissipative critical system''.
\newblock \href{https://dx.doi.org/10.1103/PhysRevA.102.052202}{Physical Review
  A {\bf 102}, 052202}~(2020).

\bibitem{goes_quantum_2020}
Bruno~O. Goes, Carlos~E. Fiore, and Gabriel~T. Landi.
\newblock ``Quantum features of entropy production in driven-dissipative
  transitions''.
\newblock \href{https://dx.doi.org/10.1103/PhysRevResearch.2.013136}{Physical
  Review Research {\bf 2}, 013136}~(2020).

\end{thebibliography}

\end{document}